# Non-Reciprocal Hyperbolic Propagation over Moving Metasurfaces

Yarden Mazor[1], and Andrea Alù[1,2,3,4*]

[1]Department of Electrical and Computer Engineering, The University of Texas at Austin, Austin, TX 78712, USA

[2]Photonics Initiative, Advanced Science Research Center, City University of New York, New York, NY 10031, USA

[3]Physics Program, Graduate Center, City University of New York, New York, NY 10026, USA

[4]Department of Electrical Engineering, City College of New York, New York, NY 10031, USA

*Hyperbolic propagation offers exciting opportunities in nanophotonics, from sub-diffraction imaging to enhanced local density of states. This transport regime is typically induced by strong modulation of conductivity, i.e., with alternating metallic and dielectric material properties. Here, we analyze a moving impedance surface, showing that suitably tailored homogeneous metasurfaces can support one-way hyperbolic propagation when in motion, adding non-reciprocity to hyperbolic propagation phenomena, and without suffering from nonlocal effects stemming from discretization or finite granularity of the surface.*

# 1 Introduction

The electrodynamics of moving media has been an active research topic for long time [1–5], highlighting various unusual properties, such as non-reciprocity and large anisotropy. Recently, with the growing interest in exotic phenomena in photonics, the interest in moving media has grown, including opportunities to induce parity-time (PT) symmetry and symmetry breaking [6,7], quantum friction [8–11] and wave instabilities [12,13]. Recently, fast-moving systems were proposed and studied, such as rapidly rotating particles [14,15] and optomechanical systems [16,17], which present promising platforms to realize these unusual effects. In a different context, hyperbolic wave propagation in metamaterials has also attracted significant interest, offering opportunities to engineer and enhance the emission of particles and molecules [18,19], for imaging and focusing [20]. To date, hyperbolic propagation has been mostly achieved using layered or wire bulk metamaterials [21], which may be accompanied by broadband non-reciprocity when a large magnetic bias is applied [22]. Hyperbolic metasurfaces, formed by alternating conductive and insulating impedance strips, enable direct and easier access to these unusual and enhanced light-matter interactions [23–25], supporting hyperbolic transport over a surface. In both approaches, however, the finite periodicity ultimately limits the exotic response by setting a limit on the cut-off wavenumber for hyperbolic propagation and introducing nonlocal effects [26]. Naturally hyperbolic materials, such as boron nitride, may provide enhanced light-matter interactions within a homogeneous bulk response, but they typically suffer from loss and are limited to narrow frequency ranges of operation [27,28]. In the following, we explore non-reciprocal hyperbolic propagation over a surface without the need of periodicity, strong modulation of the conductivity properties and of magnetic bias, but instead based on moving homogeneous surfaces. We show that motion above a certain velocity can support hyperbolic propagation with highly anisotropic and non-reciprocal responses, offering an interesting way to combine hyperbolic regimes with directional features.

## 2  Formulation

The geometry of interest, shown in the inset of figure (1a), consists of a homogeneous impedance surface moving with velocity tangent to the surface in the lab frame S. In S', the system where the surface is at rest, we use the conventional impedance boundary condition

$$\hat{\mathbf{n}} \times (\mathbf{H}'_2 - \mathbf{H}'_1) = \underline{\underline{\sigma}} \mathbf{E}'_{\tan} \quad (1)$$

where 1 (2) refers to above (below) the surface and $\underline{\underline{\sigma}}$ is the conductivity tensor. We utilize the Lorentz transformations for the electromagnetic fields [1]

$$\begin{aligned} \mathbf{E}' &= \gamma \left( \underline{\underline{\alpha}}^{-1} \mathbf{E} + c\mu_0 \underline{\underline{\beta}} \mathbf{H} \right) \\ \mathbf{H}' &= \gamma \left( -c\varepsilon_0 \underline{\underline{\beta}} \mathbf{E} + \underline{\underline{\alpha}}^{-1} \mathbf{H} \right) \end{aligned} \quad (2)$$

with $\boldsymbol{\beta} = \mathbf{v}/c$, $\beta = |\boldsymbol{\beta}|$, $\gamma = (1-\beta^2)^{-1/2}$, and the matrix operators $\underline{\underline{\alpha}}, \underline{\underline{\beta}}$ are defined in the Appendix. Upon substituting equation (2) into equation (1), after some straightforward steps we obtain the equivalent boundary condition for a tangentially moving metasurface

$$\hat{\mathbf{x}} \times (\mathbf{H}_2 - \mathbf{H}_1) = \gamma \underline{\underline{\alpha}}^{-1} \underline{\underline{\sigma}} \underline{\underline{\alpha}}^{-1} \mathbf{E}_{\tan} + \gamma (\mathbf{v} \times H_x \hat{\mathbf{x}}) + \varepsilon_0 \hat{\mathbf{n}} \cdot (\mathbf{E}_2 - \mathbf{E}_1) \mathbf{v}, \quad (3)$$

expressed in terms of the fields in the lab frame S. The right-hand side contains three electric current contributions: the first is an effective conduction current, displaying motion-induced anisotropy; the second term indicates magneto-electric coupling arising from the Lorentz force sustained by the normal magnetic field; the third term is a convection current, generated by the mechanical motion of the induced surface charge. The effective masses and distances associated with the surface structure are also altered due to the motion by a factor $\gamma$, and these second-order effects are taken into account in the effective conductivity matrix $\tilde{\underline{\underline{\sigma}}} = \gamma \underline{\underline{\alpha}}^{-1} \underline{\underline{\sigma}} \underline{\underline{\alpha}}^{-1}$. In this work, we assume the surface has an isotropic surface impedance $Z_s = 1/\sigma = jX$, with $\sigma$ being the surface conductivity ($e^{j\omega t}$ time dependence used throughout the paper), and the motion is chosen such that $\mathbf{v} = v\hat{\mathbf{z}}$ ($\boldsymbol{\beta} = \beta\hat{\mathbf{z}}$) with $v > 0$ ($\beta > 0$). We consider an inductive surface ($X > 0$) which may describe, for instance, a sheet of pristine graphene or other 2D materials in the mid-infrared range, or suitably designed metasurfaces in optics or radio-frequencies [29–32]. After substituting these assumptions into equation (3), the boundary condition used assumes the form

$$\hat{\mathbf{x}} \times (\mathbf{H}_1 - \mathbf{H}_2) = \tilde{\underline{\underline{\sigma}}} \mathbf{E}_{\tan} + \sigma\gamma \left( v\hat{\mathbf{z}} \times \mu_0 \mathbf{H}_x \right) + (v\hat{\mathbf{z}})\hat{\mathbf{n}} \cdot (\varepsilon_0 \mathbf{E}_1 - \varepsilon_0 \mathbf{E}_2), \tag{4}$$

with the effective conductivity now expressed in the simple form $\tilde{\underline{\underline{\sigma}}} = \sigma \begin{pmatrix} \gamma & 0 \\ 0 & \gamma^{-1} \end{pmatrix}$.

## 3 Quasi-TM one-way hyperbolic modes

In S', only transverse-magnetic (TM) modes are supported, as expected for homogeneous inductive impedance surfaces [33,34]. Due to the anisotropy induced by the motion, the surface waves propagating in S will no longer be pure-TM [30] when considering propagation into various angles, but since they are obtained from the transformation of pure-TM waves in S', and for moderate speeds are still dominated by their TM component, we shall term them quasi-TM (qTM). The electromagnetic fields associated with the surface waves have the form $e^{-j\mathbf{k}_t \cdot [y,z]} e^{-\alpha|x|}$, with in-plane wave vector $\mathbf{k}_t$ and confinement coefficient $\alpha$. Their dispersion relation in S is obtained by substituting the surface-wave fields into boundary condition (4), which results in

$$\alpha = \sqrt{\delta}\left(k_0 - \beta k_{tz}\right), \tag{5}$$

with $\delta = (2\gamma X / \eta_0)^2$. Combining this with the free-space dispersion ($k_{tz}^2 + k_{ty}^2 = \alpha^2 + k_0^2$) gives the in-plane dispersion iso-frequency contour in S

$$[1-\delta\beta^2]\left(\frac{k_{tz}}{k_0} + \frac{\beta\delta}{1-\delta\beta^2}\right)^2 + \left(\frac{k_{ty}}{k_0}\right)^2 = 1+\delta+\frac{\beta^2\delta^2}{1-\delta\beta^2}. \tag{6}$$

The dispersion is clearly non-reciprocal, due to the odd $k_{tz}$ term, induced by the linear motion that breaks time-reversal symmetry. Its expression yields an ellipse ($\delta\beta^2 < 1$) or hyperbola ($\delta\beta^2 > 1$), as shown in Fig. (1a) for varying values of $\beta$. The threshold of $\beta$ for which the topological transition occurs is

$$\beta_{th,TM} = \frac{v_{p,TM}}{c}, \tag{7}$$

where $v_{p,TM} = c\left(1+4X^2/\eta_0^2\right)^{-1/2}$ is the phase velocity of the TM surface waves on the stationary surface. Since $\alpha$ in equation (5) must be positive for the wave not to diverge, we additionally obtain $\beta k_{tz} < k_0$. When $\beta > \beta_{th,TM}$ this inequality forbids one branch of the hyperbola in (6), yielding a one-way hyperbolic dispersion contour, as seen in Fig. 1, where the slope of the hyperbola asymptote is $\gamma\sqrt{v^2/v_{p,TM}^2 - 1}$.

In the qTM regime, due to the nature of the fields the (normal electric field, very weak normal magnetic field) the convection current is the dominant term compared to the Lorentz current. Therefore, in this regime we may define an effective qTM conductivity tensor

$$\underline{\underline{\tilde{\sigma}}}\mathbf{E}_{\tan} + (v\hat{\mathbf{z}})\hat{\mathbf{n}} \cdot (\varepsilon_0 \mathbf{E}_1 - \varepsilon_0 \mathbf{E}_2) = \begin{pmatrix} \sigma\gamma & 0 \\ 2jvk_{ty}\alpha^{-1} & \sigma\gamma^{-1} + 2jvk_{tz}\alpha^{-1} \end{pmatrix} \mathbf{E}_{\tan} \tag{8}$$

which captures the propagation properties in this system: hyperbolic propagation is associated with a change of sign of the first diagonal term of (8) for large enough values of $v$, and non-reciprocity arises from the odd-dependence in $k_{tz}$.

We validated our analysis calculating with a full-wave electromagnetic solver (COMSOL), the fields induced on a finite segment of moving impedance by a source on the left side of the strip, and the motion was modeled using the effective boundary condition in equation (4). Figure (1b) shows the a snapshot of the longitudinal electric field $\text{Re}\{E_z\}$ when the surface is static and $X = 5\eta_0$ ($\eta_0 = \sqrt{\mu_0/\varepsilon_0}$) when exciting from the $z = 0$ by an aperture field distribution corresponding to the $+\hat{\mathbf{z}}$ propagating wave. Here, the supported surface wave has the expected wavenumber $k_{tz} \sim 10k_0$. For $\beta = 0.05$ [Figure (1c)] non-reciprocity, as evident from the different wavelength for forward and backward waves. The extracted wave-numbers, after Fourier transforming the fields (shown in figure (1d)), are consistent with our analytical dispersion, $k_{tz,forward} \approx 6.7k_0$ and $k_{tz,backward} \approx 20.15k_0$. Nonreciprocal propagation of surface waves was also discussed in [35], where nonreciprocity was induced by a drift current driven over graphene and in [36] through a metallic slab; here mechanical motion effectively replaces the current bias.

## 4 Quasi-TE one-way hyperbolic modes

Interestingly, the surface motion enables transverse-electric (TE) surface modes, which are forbidden along inductive impedance surfaces at rest. The dispersion equation for quasi-TE (qTE) surface waves can be obtained in the same way as qTM, and it reads

$$\alpha_{TE} = -\frac{\eta_0}{2X}\gamma(k_0 - \beta k_t \cos\varphi). \tag{9}$$

For low speeds, $\alpha_{TE} > 0$ can be satisfied only for capacitive surfaces $X < 0$. However, $X$ can be positive in (9) when the term in brackets becomes negative. In this case, while the surface at rest is inductive, the motion enables TE propagation. Interestingly, these waves can only have non-reciprocal hyperbolic dispersion and, using Eq. (9), we find that these modes are supported for velocities satisfying $\beta > \beta_{th,TE} = \left(1 + (\eta_0/2X)^2\right)^{-1/2}$. For large surface inductance values, like those examined in figure (1), i.e., far from resonance, this value is close to 1, implying fast required speeds. However, this requirement can be relaxed using lower inductance values, for metasurfaces closer to resonance, enabling unique propagation features of both qTE and qTM modes. The effective boundary condition in (4) shows that for qTE propagation the conduction and Lorentz current currents are dominant comparing to the convection current, which defines the equivalent qTE conductivity

$$\underline{\underline{\tilde{\sigma}}}\mathbf{E}_{tan} + \underline{\underline{\tilde{\sigma}}}(v\hat{\mathbf{z}} \times \mu_0\mathbf{H}_x) = \begin{pmatrix} \gamma\sigma(1 - \beta k_{tz}/k_0) & \gamma\sigma\beta k_{ty}/k_0 \\ 0 & \sigma\gamma^{-1} \end{pmatrix}\mathbf{E}_{tan}. \tag{10}$$

In figure (2a) we show the dispersion of qTE waves on an inductive surface for increasing velocity, with $X = \eta_0/20$, yielding a threshold value of $\beta_{th,TE} \simeq 0.1$. As the velocity increases the dispersion hyperbola become wider. Figure (2b) maps various propagation regimes for qTE and qTM modes vs. $X, \beta$. The black curves represent the threshold values $\beta_{th,TE}, \beta_{th,TM}$. For low velocities only anisotropic qTM propagation is possible, but as the speed increases additional regimes arise: high-inductance surfaces allow one-way hyperbolic qTM modes, whereas low-inductance surfaces allow hyperbolic qTE modes. For high velocities, both hyperbolic regimes are possible. Figure (2c) shows the emission of a magnetic 2D dipole $\mathbf{m} = m\hat{\mathbf{y}}$ exciting a TE mode for $\beta = 0.15 > \beta_{th,TE}$, and one-way efficient emission takes place. Fig. (2d) shows qTM excitation in the elliptic regime, yielding anisotropic nonreciprocal propagation. Fig. (2e) shows qTM excitation in the hyperbolic regime for the same surface at a faster speed; one-way hyperbolic emission is visible, with enhanced emission rates and stronger spatial localization. Fig. (2f) shows excitation of the surface in Fig. (2c) by a magnetic dipole, inducing one-way qTE hyperbolic surface waves. Both hyperbolic regimes display high intensity of the excited waves in the directions parallel to the hyperbola asymptotes, which leads to the expected light-matter interaction enhancement, here uniquely combined with strong non-reciprocal response.

Close examination reveals that qTE propagation over moving inductive surfaces arises from TE modes excited at negative frequencies in S'. Heuristically, Eq. (9) shows that in S' TE modes are supported on inductive surfaces ($X > 0$) if $\omega < 0$. Negative frequencies in S' can be Doppler shifted to positive in S for sufficiently large velocities, allowing access to these modes. Coupling of radiation processes with negative frequency waves using motion was studied in [12,37].

## 5   Effect of losses

When losses are considered, we expect waves with different wavenumbers to have different decay constants. To incorporate losses in our model, we let the surface impedance obtain a complex value - $Z_s = jX = -X_I + jX_R$. In this scenario, $\alpha, k_t$ are also complex-valued, $\alpha = \alpha_R + j\alpha_I$ , $k_t = k_{tR} + jk_{tI}$, and for simplicity we assume $\hat{\mathbf{z}}$ propagation. considering qTM propagation, and substituting the complex valued parameters into the boundary condition in equation (4), we get

$$\alpha_R + j\alpha_I = 2\gamma(X_R + jX_I)(k_0 - \beta k_{tR} - j\beta k_{tI}) \tag{11}$$

And an example for the asymmetric attenuation is shown in figure (3). We see that as the speed of the surface increases, the separation in the attenuation coefficients becomes larger, in conjunction with the real parts of $k_z$. This can be attributed to the fact that larger real parts of the wavenumber imply stronger confinement of the fields, which lead to enhanced absorption, larger imaginary parts of $k_t$, and asymmetric propagation distances.

# 6 Conclusions

We have shown that moving metasurfaces enable a unique regime of non-reciprocal hyperbolic wave propagation, supporting the insurgence of TM and TE surface modes coupled over the same surface, which enable the directional emission of localized electric and magnetic emitters over a surface with strongly localized enhanced light-matter interactions. While the required speeds may be impractical in some scenarios, one may consider alternative systems to qualitatively demonstrate some of these effects, such as rotating surfaces [14,15], or space-time modulated surfaces that effectively mimic motion [38–40]. We are currently exploring these opportunities.

# 7 Appendix

To compactly define the electromagnetic field Lorentz transformations we use matrix operators defined in [1], listed here for the sake of completeness. The operator $\underline{\underline{\boldsymbol{\beta}}}$ represents the $\boldsymbol{\beta}\times$ operation

$$\underline{\underline{\boldsymbol{\beta}}} = \boldsymbol{\beta} \times \underline{\underline{\mathbf{I}}} = \begin{pmatrix} 0 & -\beta_z & \beta_y \\ \beta_z & 0 & -\beta_x \\ -\beta_y & \beta_x & 0 \end{pmatrix} \tag{12}$$

And $\underline{\underline{\boldsymbol{\alpha}}}$ is defined as

$$\underline{\underline{\boldsymbol{\alpha}}} = \underline{\underline{\mathbf{I}}} + (\gamma - 1)\frac{\boldsymbol{\beta}\boldsymbol{\beta}}{\beta^2} \tag{13}$$

Where $\underline{\underline{\mathbf{I}}}$ is the 3x3 unit matrix, and $\boldsymbol{\beta}\boldsymbol{\beta}$ is the external product.

[21] A. Poddubny, I. Iorsh, P. Belov, and Y. Kivshar, Nat. Photonics **7**, 958 (2013).

[22] A. Leviyev, B. Stein, A. Christofi, T. Galfsky, H. Krishnamoorthy, I. L. Kuskovsky, V. Menon, and A. B. Khanikaev, APL Photonics **2**, 076103 (2017).

[23] J. S. Gomez-Diaz, M. Tymchenko, and A. Alù, Opt. Mater. Express **5**, 2313 (2015).

[24] J. S. Gomez-Diaz, M. Tymchenko, and A. Alù, Phys. Rev. Lett. **114**, 233901 (2015).

[25] J. S. Gomez-Diaz and A. Alù, ACS Photonics **3**, 2211 (2016).

[26] D. Correas-Serrano, J. S. Gomez-Diaz, M. Tymchenko, and A. Alù, Opt. Express **23**, 29434 (2015).

[27] S. Dai, Q. Ma, S. Zhu, M. K. Liu, T. Andersen, Z. Fei, M. Goldflam, M. Wagner, K. Watanabe, T. Taniguchi, M. Thiemens, F. Keilmann, G. Janssen, P. Jarillo-Herrero, M. M. Fogler, and D. N. Basov, Nat. Nanotechnol. **10**, 682 (2015).

[28] S. Dai, M. Tymchenko, Y. Yang, Q. Ma, M. Pita-Vidal, K. Watanabe, T. Taniguchi, P. Jarillo-Herrero, M. M. Fogler, A. Alù, and D. N. Basov, Adv. Mater. **30**, 1706358 (2018).

[29] K. Achouri, M. A. Salem, and C. Caloz, IEEE Trans. Antennas Propag. **63**, 2977 (2015).

[30] H. J. Bilow, IEEE Trans. Antennas Propag. **51**, 2788 (2003).

[31] A. Epstein and G. V. Eleftheriades, IEEE Trans. Antennas Propag. **64**, 3880 (2016).

[32] S. B. Glybovski, S. A. Tretyakov, P. A. Belov, Y. S. Kivshar, and C. R. Simovski, Phys. Rep. **634**, 1 (2016).12

**Figures**

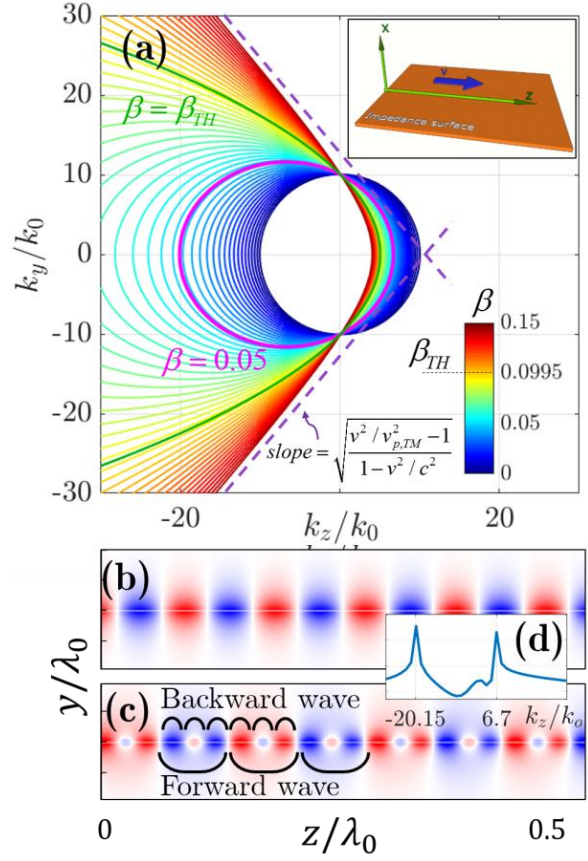

**Figure 1.** (a) Isofrequency contours for the normalized wavenumber. The value of $\beta$ is color coded. A topological transition from elliptical to hyperbolic is noticed around the threshold value $\beta_{TH} \approx 0.1$ (green thick line), also highlighted in the color bar. The surface inductance is $X = 5\eta_0$. The inset shows the geometry of interest. (b) $E_z$ distribution when the surface is stationary. (c) $E_z$ distribution when $\beta = 0.05$, corresponding to the magenta curve in panel (a). (d) Fourier transform of $E_z$ presented in (c). The dominant wavenumbers are labeled.

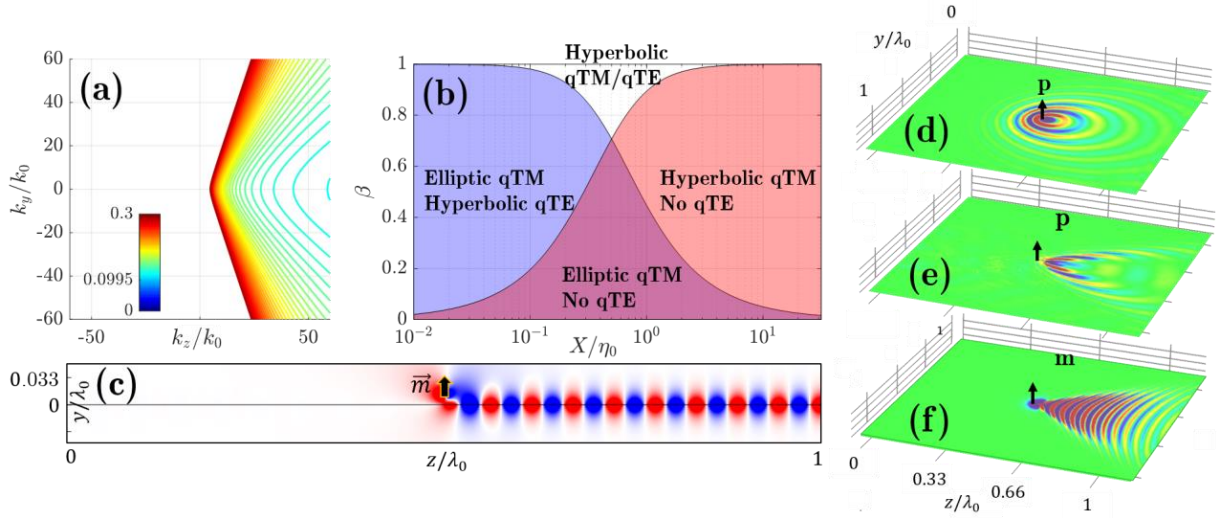

**Figure 2**. (a) Dispersion of the qTE modes on a moving inductive surface, color coded according to the value of $\beta$, for $X/\eta_0 = 1/20$. (b) Map of the possible guiding regimes of qTM and qTE modes over a moving inductive surface. (c) Excitation of one-way TE surface wave on a moving impedance surface by a 2D magnetic dipole with $\beta = 0.15$. (d) Excitation of qTM waves in the elliptic regime, $X/\eta_0 = 5, \beta = 0.05$ (purple region in (b)). (e) Excitation of qTM surface waves in the hyperbolic regime, $X/\eta_0 = 5, \beta = 0.15$ ( orange region). (f) Excitation of qTE surface waves in the hyperbolic regime, same parameters as (c) (purple region).

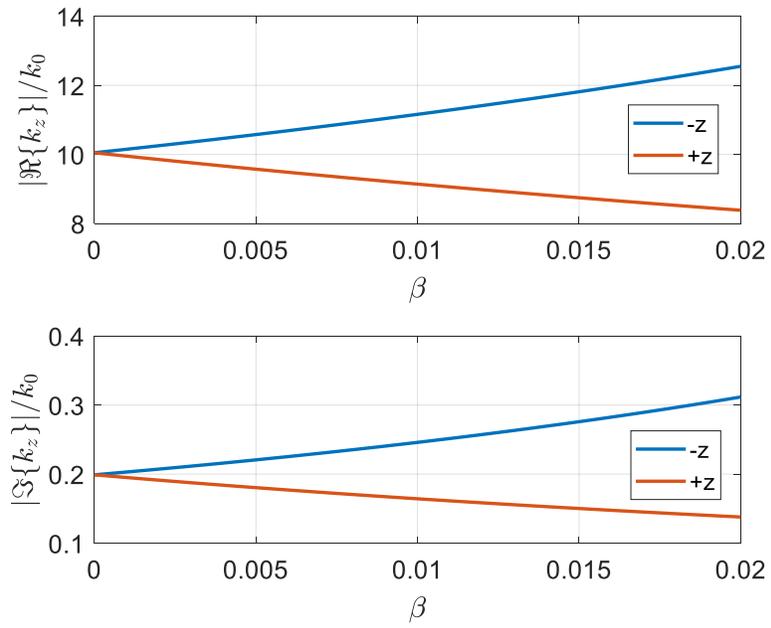

**Figure 3**. *Real and imaginary parts of the longitudinal wavenumber $k_z$ for propagation on lossy impedance surface. The conductivity parameters chosen here are $X_R = 5\eta_0$ and $X_I = 0.1\eta_0$.*